\documentclass[twocolumn,showpacs,preprintnumbers,amsmath,amssymb]{revtex4}

\usepackage{graphicx}
\usepackage{dcolumn}
\usepackage{bm}

\begin{document}

\title{Unveiling New Magnetic Phases of Undoped and Doped Manganites}

\author{Takashi Hotta$^1$, Mohammad Moraghebi$^2$, Adrian Feiguin$^3$,
Adriana Moreo$^2$, Seiji Yunoki$^4$, and Elbio Dagotto$^2$}

\affiliation{
$^1$Advanced Science Research Center,
Japan Atomic Energy Research Institute,
Tokai, Ibaraki 319-1195, Japan \\ 
$^2$National High Magnetic Field Laboratory,
Florida State University, Tallahassee, Florida 32306 \\
$^3$Department of Physics and Astronomy,
University of California at Irvine, CA 92697 \\
$^4$International School for Advanced Studies (SISSA),
via Beirut 4, 34014 Trieste, Italy
}

\date{\today}

\begin{abstract}
Novel ground-state spin structures in undoped and lightly-doped manganites are
here investigated based on the orbital-degenerate double-exchange model,
by using mean-field and numerical techniques.
In undoped manganites, a new antiferromagnetic (AFM) state, called
the E-type phase, is found adjacent in parameter space 
to the A-type AFM phase.
Its structure is in agreement with recent experimental results.
This insulating E-AFM state is also competing with a ferromagnetic metallic
phase as well, suggesting that large magneto-resistant effects could
exist even in undoped Mn oxides.
For doped layered manganites, the phase diagram includes another new
AFM phase of the ${\rm C_x E_{1-x}}$-type.
Experimental signatures of the new phases are discussed.
\end{abstract}



\maketitle


In the recent decade, the study of manganites -- materials that show a
remarkable Colossal Magneto-Resistance (CMR) \cite{Tokura} -- 
has been one of the most important areas of research in condensed 
matter \cite{Dagotto1}.
This CMR effect occurs when the manganite ground-state changes from
insulating to ferromagnetic (FM) metallic after a small magnetic field
is applied.
Based on the concept of two-phase competition \cite{Dagotto1},
the CMR behavior has been successfully qualitatively reproduced in
computational simulations employing resistor-network models \cite{CMR}.
However, more work remains to be done to fully understand Mn-oxides,
both regarding their unusual magneto-transport properties
and the nature of the many competing phases.

The appearance of the FM metallic phase in manganites is usually
rationalized by the so-called double-exchange (DE) mechanism,
based on a strong Hund coupling between mobile $e_{\rm g}$ electrons
and localized $t_{\rm 2g}$ spins.
On the other hand, the insulating phase in manganites occurs
due to the coupling between degenerate $e_{\rm g}$ electrons and
Jahn-Teller (JT) distortions of the MnO$_6$ octahedra,
leading to the various types of charge and/or orbital orders observed
experimentally \cite{Dagotto1}.

The parent compound of CMR manganites is undoped RMnO$_3$,
where R denotes rare earth ions.
For R=La, as is well-known, the A-type antiferromagnetic (AFM) phase
appears, with the C-type ordering of $(3x^2$$-$$r^2)$- and
$(3y^2$$-$$r^2)$-orbitals \cite{LaMnO3}.
By substituting La by alkaline earth ions such as Sr and Ca, holes are doped
into the $e_{\rm g}$-electron band and due to the DE mechanism,
the FM metallic phase appears, with its concomitant CMR effect.
Most of the discussion in manganites has centered on the many phases
induced by doping with holes the A-type AFM state, at different values
of their bandwidths. In this framework, it is implicitly assumed
that the undoped material is always in an A-type state.
However, quite recently, a new AFM phase has been reported as the ground-state 
in the undoped limit for R=Ho \cite{Munoz,Kimura}.
This phase is here called the ``E-type'' spin structure
following the standard notation in this context.
It is surprising that a new phase can be still found even in the
undoped material, previously considered to be well understood.
In addition, the nature of the states obtained by lightly doping this E-phase
is totally unknown, and new phenomena may be unveiled experimentally in
the near future. Overall, {\it this opens an exciting new branch of 
investigations in manganites since novel phases appear to be hidden
in the vast parameter space of these compounds}. A clear example
has been recently provided by the prediction of a FM charge-ordered (CO)
phase at x=1/2 \cite{Hotta4,Yunoki},
which may have been found experimentally already \cite{Mathur}.

In this Letter, based on the orbital-degenerate DE model coupled with
JT distortions, the ground state properties of undoped manganites
are analyzed by using mean-field (MF) calculations
and Monte-Carlo (MC) simulations.
In our phase diagram at x=0, the E-AFM phase is found to exist in a $wide$
region of parameter space, adjacent to the A-AFM phase in agreement
with the experimental results.
The E-AFM phase is robust when the dimensionality and/or electron-phonon
coupling are modified, and its strength originates in
the intrinsic nature of orbital-degenerate DE systems.
In the phase diagram and at small electron-phonon coupling
the E-AFM insulating phase is located just
next to a FM metallic phase \cite{comment3},
suggesting that CMR effects could be found even at x=0.
Light-hole doping x of the E-type phase is also discussed and another
novel magnetic phase, defined as the ``${\rm C_xE_{1-x}}$'' phase, is found.
The ubiquitous phase-separation tendencies observed when insulating and
metallic phases compete is also expected near the 
FM--${\rm C_xE_{1-x}}$ boundary.

The Hamiltonian studied in this paper is
\begin{eqnarray}
  H &=& -\sum_{{\bf ia}\gamma \gamma'\sigma}
  t^{\bf a}_{\gamma \gamma'} d_{{\bf i} \gamma \sigma}^{\dag}
  d_{{\bf i+a} \gamma' \sigma}
  -J_{\rm H} \sum_{\bf i}
  {\bf s}_{\bf i} \cdot {\bf S}_{\bf j} \nonumber \\
  &+& J_{\rm AF} \sum_{\langle {\bf i,j} \rangle}
  {\bf S}_{\bf i} \cdot {\bf S}_{\bf j}
  + \lambda \sum_{\bf i}
  (Q_{1{\bf i}}\rho_{\bf i} + Q_{2{\bf i}}\tau_{{\rm x}{\bf i}} 
  +Q_{3{\bf i}}\tau_{{\rm z}{\bf i}}) \nonumber \\
  &+& (1/2) \sum_{\bf i} (\beta Q_{1{\bf i}}^2
  +Q_{2{\bf i}}^2+Q_{3{\bf i}}^2),
\end{eqnarray}
where $d_{{\bf i}{\rm a}\sigma}$ ($d_{{\bf i}{\rm b}\sigma}$)
annihilates an $e_{\rm g}$-electron with spin $\sigma$
in the $d_{x^2-y^2}$ ($d_{3z^2-r^2}$) orbital at site ${\bf i}$,
and ${\bf a}$ is the vector connecting nearest-neighbor (NN) sites.
The first term is the NN hopping of $e_{\rm g}$ electrons
with amplitude $t^{\bf a}_{\gamma \gamma'}$ between
$\gamma$- and $\gamma'$-orbitals along the ${\bf a}$-direction:
$t^{\bf x}_{\rm aa}$=$-\sqrt{3}t^{\bf x}_{\rm ab}$=
$-\sqrt{3}t^{\bf x}_{\rm ba}$=$3t^{\bf x}_{\rm bb}$=$t$
for ${\bf a}$=${\bf x}$,
$t^{\bf y}_{\rm aa}$=$\sqrt{3}t^{\bf y}_{\rm ab}$=
$\sqrt{3}t^{\bf y}_{\rm ba}$=$3t^{\bf y}_{\rm bb}$=$t$
for ${\bf a}$=${\bf y}$,
and $t^{\bf z}_{\rm bb}$=$4t/3$ with
$t^{\bf z}_{\rm aa}$=$t^{\bf z}_{\rm ab}$=$t^{\bf z}_{\rm ba}$=0
for ${\bf a}$=${\bf z}$.
Hereafter, $t$ is taken as the energy unit.
In the second term, the Hund coupling $J_{\rm H}$($>$0)
links $e_{\rm g}$ electrons with spin
${\bf s}_{\bf i}$=
$\sum_{\gamma\alpha\beta}d^{\dag}_{{\bf i}\gamma\alpha}
\bm{\sigma}_{\alpha\beta} d_{{\bf i}\gamma\beta}$
($\bm{\sigma}$=Pauli matrices)
with the localized $t_{\rm 2g}$ spin ${\bf S}_{\bf i}$
assumed classical with $|{\bf S}_{\bf i}|$=1.
$J_{\rm H}$ is here considered as infinite or very large.
The third term is the AFM coupling $J_{\rm AF}$
between NN $t_{\rm 2g}$ spins.
The fourth term couples $e_{\rm g}$ electrons
and MnO$_6$ octahedra distortions \cite{note:Coulomb1},
$\lambda$ is a dimensionless coupling constant,
$Q_{1{\bf i}}$ is the breathing-mode distortion,
$Q_{2{\bf i}}$ and $Q_{3{\bf i}}$ are, respectively, 
$(x^2$$-$$y^2)$- and $(3z^2$$-$$r^2)$-type JT-mode distortions,
$\rho_{\bf i}$=
$\sum_{\gamma,\sigma}
d_{{\bf i}\gamma\sigma}^{\dag}d_{{\bf i}\gamma\sigma}$,
$\tau_{{\rm x}{\bf i}}$=
$\sum_{\sigma}(d_{{\bf i}{\rm a}\sigma}^{\dag}d_{{\bf i}{\rm b}\sigma}$
+$d_{{\bf i}{\rm b}\sigma}^{\dag}d_{{\bf i}{\rm a}\sigma})$,
and
$\tau_{{\rm z}{\bf i}}$=
$\sum_{\sigma}(d_{{\bf i} a\sigma}^{\dag}d_{{\bf i}a\sigma}$
$-$$d_{{\bf i} b\sigma}^{\dag}d_{{\bf i}b\sigma})$.
The fifth term is the usual quadratic potential for adiabatic distortions
and $\beta$ is the spring-constants ratio for breathing- and JT-modes.
In actual manganites, $\beta$$\approx$2.
In undoped manganites, all oxygens are shared by adjacent MnO$_6$
octahedra and the distortions are not independent,
suggesting that the cooperative effect is even more important than
for the doped case x$>$0.
To consider this cooperation, here oxygen ion displacements are directly
optimized \cite{note:oxygen}.

\begin{figure}[t]
\includegraphics[width=1.0\linewidth]{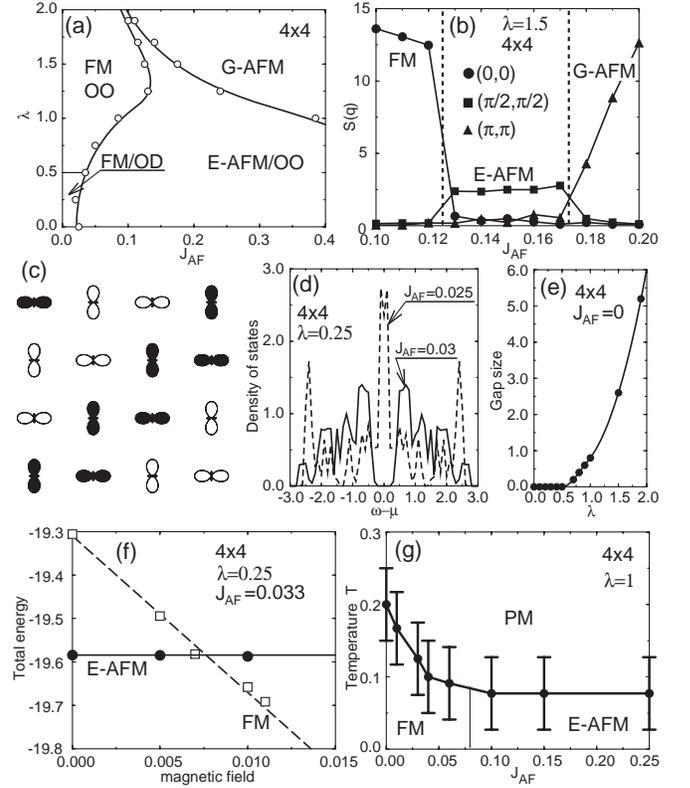}
\caption{
(a) Phase diagram obtained using the 4$\times$4 lattice.
Open circles denote MC results and solid lines
are obtained with MF calculations.
(b) Spin correlation $S({\bf q})$ vs. $J_{\rm AF}$ at $\lambda$=1.5.
(c) Schematic view of the spin and orbital structure of the E-AFM phase.
Solid and open symbols denote orbitals with up- and down-spins,
respectively. Note the presence of zigzag chains for each spin orientation.
(d) Density of states (DOS) obtained by MC simulations at $\lambda$=0.25.
Solid and dashed curves denote the results for E- and FM-phases.
(e) Gap at the Fermi level for the FM phase vs. $\lambda$.
The magnitude is evaluated from the DOS obtained by MC simulations.
(f) Total energy vs. magnetic field for E-AFM and FM phases.
(g) Phase diagram in the ($J_{\rm AF}$, $T$) plane at x=0 and $\lambda$=1,
using a 4$\times$4 lattice. Estimations of the Curie and N\'eel temperatures
are evaluated using MC simulations (see text).
}
\end{figure}

Let us first describe our two-dimensional (2D) results,
since the essential physics behind the stabilization of the E-type phase
can be grasped by using MC simulations with relatively short CPU times.
The phase diagram on a 4$\times$4 lattice is shown in Fig.~1(a).
The continuous curves are obtained by 
comparing the energies of the competing phases in the MF calculations,
while the circles are obtained by monitoring the nature of the dominant
spin correlation $S({\bf q})$ in MC simulations.
A typical result for $S({\bf q})$ is shown in Fig.~1(b).
A new regime characterized by ${\bf q}$=$(\pi/2,\pi/2)$
is clearly observed between the FM and G-AFM phases.
The good agreement between MF and MC results in the region of
interest shows the high accuracy of the present
MF calculations for manganites \cite{Hotta3}.

In Fig.~1(c), the spin and orbital structure of the novel intermediate
phase (E-phase) is shown.
Along the zigzag chains, $t_{\rm 2g}$ spins order ferromagnetically,
but they are antiparallel perpendicular to the zigzag direction.
In the three-dimensional (3D) case, the MF study shows that the pattern Fig.~1(c) 
just $stacks$ along the $z$-axis, while the spin directions are
reversed from plane to plane.
Note that the orbital structure is the same as that
of the A-AFM phase, namely, the staggered pattern of ($3x^2$$-$$r^2$)-
and ($3y^2$$-$$r^2$)-like orbitals.
Our investigations show that the E-phase is robust 
at weak- and intermediate-$\lambda$, but 
for $\lambda$$>$1.5, the E-type regime narrows.

A surprising aspect of our results is that the E-type spin arrangement
is the ground-state for a $wide$ range of $J_{\rm AF}$, even at $\lambda$=0,
indicating that the coupling with JT phonons is $not$ a necessary condition
for its stabilization.
This is in sharp contrast to the case of the A-AFM phase.
To understand this point, it is instructive to study the
$e_{\rm g}$-electronic structure of the zigzag FM chains that appear
in the E-phase spin arrangement.
Taking $J_{\rm H}$ as infinity for simplicity,
the $e_{\rm g}$ electrons move only along the zigzag FM chain,
and cannot hop to the adjacent FM chains.
The dispersion energy for $e_{\rm g}$ electrons in this zigzag FM chain
is given by $\varepsilon_{k}$=$(2/3)(\cos k$$\pm$$\sqrt{\cos^2 k+3})$
and $(2/3)(-\cos k$$\pm$$\sqrt{\cos^2 k+3})$,
indicating that there appears a large {\it band-gap} equal to $4t/3$
at half-filling.
In fact, even on the 4$\times$4 cluster,
there is a clear gap of the order of $t$ for the E-phase
in the density of states (DOS) (Fig.~1(d)).
Since $t^{\bf x}_{\mu\nu}$=$-t^{\bf y}_{\mu\nu}$ for $\mu$$\ne$$\nu$,
the sign in the hopping amplitude changes periodically
in the zigzag situation,
leading to a periodic potential for $e_{\rm g}$ electrons and
its concomitant band-insulator nature.
In other words, {\it the E-type phase is stable due to the 
zigzag geometry of the FM chains that induce a band-insulator}
\cite{note:CE}.

Another related interesting point is the orbital structure of the FM phase.
In the strong-coupling region, an orbitally ordered (OO) state appears,
essentially with the same pattern as that observed in the E-AFM phase
discussed above.
All MnO$_6$ octahedra are distorted at x=0 and the cooperative effect
is essential to determine the OO pattern for large $\lambda$,
irrespective of the spin structure.
However, when $\lambda$ decreases OO $disappears$ and
an orbital disordered (OD) new phase is observed.
This is a metallic phase according to the DOS in Fig.~1(d).
The OO-OD transition is monitored by the gap size
at the Fermi level in the DOS (Fig.~1(e)).
The transition observed using a 4$\times$4 cluster is robust,
although the actual critical value may change in larger systems \cite{Hotta7}.

Note that the OD/FM phase is next to the insulating E-AFM phase
for $\lambda$$\alt$0.5 (the E-phase is insulating with a DOS gap both
at small and large $\lambda$).
Such a result is new in the study of undoped manganites.
Since the competition between FM metallic and insulating phases
is at the heart of the CMR phenomena,
by tuning experimentally the lattice parameters in RMnO$_3$
it may be possible to observe the magnetic-field induced metal-insulator
transition {\it even in undoped manganites}.
In fact, small magnetic fields of a few Teslas indeed induce
a phase transition between E- and FM phases (see Fig.~1(f)),
with a concomitant jump in the magnetization.
 
Let us discuss the effect of thermal fluctuations.
For this purpose, both $T_{\rm C}$ (Curie temperature) and $T_{\rm N}$
(N\'eel temperature) are estimated from MC simulations
in 4$\times$4 clusters \cite{note:TcandTN} (Fig.~1(g)).
Our estimated $T_{\rm C}$ is found to decrease with increasing
$J_{\rm AF}$, and in the E-AFM phase, $T_{\rm N}$ becomes almost constant.
The E-phase can be disordered by thermal fluctuations faster upon
heating than the FM phase.
Provided that our estimated $T_{\rm C}$ in 2D system is related to
$T_{\rm N}$ with the A-AFM phase of 3D lattices,
Fig.~1(g) mimics well the experimental results for the changes
of $T_{\rm N}$ by R-ion substitution in RMnO$_3$ \cite{Kimura}. 

\begin{figure}
\includegraphics[width=1.0\linewidth]{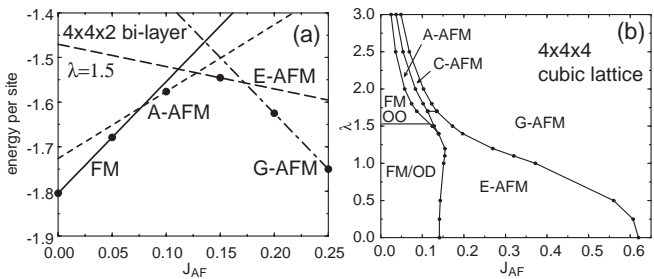}
\caption{
(a) Energies of the FM, A-AFM, E-AFM, C-AFM, and G-AFM phases
on 4$\times$4$\times$2 bi-layer lattices ($\lambda$=1.5).
Solid circles indicate the results of optimizations, while lines
denote the MF results.
(b) Phase diagram for the 4$\times$4$\times$4 cubic lattice.
All solid lines emerge from MF calculations.
}
\end{figure}

Now let us turn our attention to the effect of dimensionality.
In Fig.~2(a), using bi-layer 4$\times$4$\times$2 clusters,
the ground-state energy per site is depicted vs. $J_{\rm AF}$
at $\lambda$=1.5 both for numerical optimization \cite{note:opt}
and MF calculations, with a good agreement between them.
Figure 2(b) is the phase diagram on a 4$\times$4$\times$4 cubic lattice,
obtained only by the MF approximation, since MC calculations in this
case are too CPU time consuming (previous MF-MC comparisons
suggest that this MF phase diagram is accurate).
In the strong-coupling region, there occurs a chain of transitions
from FM$\rightarrow$A-AFM$\rightarrow$C-AFM$\rightarrow$G-AFM
phases \cite{note:FMOO}, already obtained in 2$\times$2$\times$2
calculations \cite{LaMnO3}.
The present result shows that size effects are small in undoped
strongly-coupled manganites, which is intuitively reasonable. 
Note that near $\lambda$$\sim$1.6, a realistic value for manganites,
the A-AFM phase is adjacent to the E-type state.
This region could correspond to the actual situation observed
in experiments for RMnO$_3$:
When the ionic radius of the R-site decreases, 
$T_{\rm N}$ of the A-AFM phase decreases as well,
and eventually the E-AFM phase is stabilized for R=Ho \cite{Kimura}.
In the weak-coupling region, the E-type phase is stable
in a wide range of $J_{\rm AF}$, as in the 2D calculation.

\begin{figure}
\includegraphics[width=1.0\linewidth]{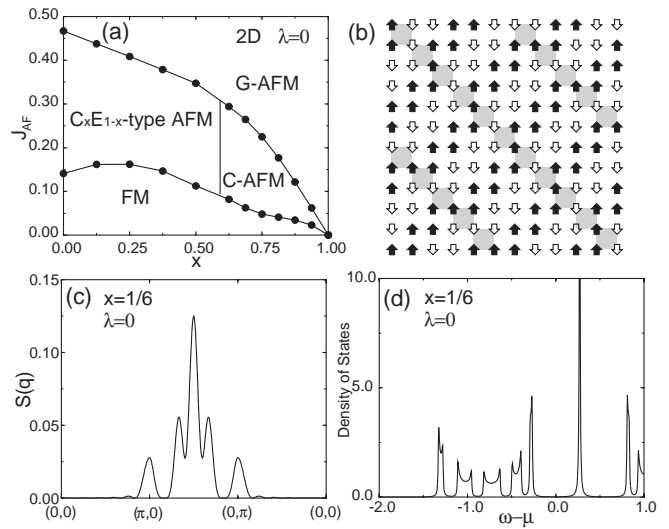}
\caption{
(a) Phase diagram in the (x,$J_{\rm AF}$) plane for layered manganites
at $\lambda$=0 obtained by analytic calculations.
(b) Schematic view for spin structure of the C$_{\rm x}$E$_{\rm 1-x}$-type phase
at x=1/6. Hatched squares denote hole-rich C-type regions.
(c) $S({\bf q})$ and (d) the DOS in the C$_{1/6}$E$_{5/6}$-type phase.
}
\end{figure}

Consider now the very interesting effect of light hole doping
on the E-phase.
Hole doping will be here studied in the weak coupling limit,
since the E-type phase is well understood at $\lambda$=0.
Figure 3(a) shows the ground-state phase diagram in the
(x, $J_{\rm AF}$) plane, obtained using analytic calculations
on 2D lattices at $\lambda$=0.
A remarkable feature of this phase diagram 
is the appearance of the novel C$_{\rm x}$E$_{\rm 1-x}$ phase,
composed of long-period zigzag FM chains, antiferromagnetically
coupled to each other (see Fig.~3(b) for x=1/6). 
As the doping fraction grows, it is expected that the 
previously reported states at intermediate- and high-doping will
eventually be reached \cite{Hotta2,Hotta4}

Note that the curvature of the boundary between
FM and C$_{\rm x}$E$_{\rm 1-x}$ phases is found to be $negative$
for 0$\alt$x$\alt$0.5, indicating that phase separation occurs
between those two phases.
Here it is again stressed that the FM phase is metallic, while
C$_{\rm x}$E$_{\rm 1-x}$ is insulating.
The latter can be considered as a microscopic phase separated state,
since the C- and E-type structure are mixed at the length scale
of a lattice constant.
This phase is expected to be made unstable easily near the phase
boundary region, and to turn into a phase-separated state with
a mixture of metallic FM and insulating C$_{\rm x}$E$_{\rm 1-x}$
clusters, which may induce CMR effects.

In the C$_{\rm x}$E$_{\rm 1-x}$ phase, $e_{\rm g}$ holes tend to
localize in the C-type region (hatched squares in Fig.~3(b)),
i.e., in the straight segment portion of the zigzag FM chain,
and as a consequence charge-ordering of the stripe form
is induced even in the $\lambda$=0 limit \cite{note:Coulomb2}. 
This causes incommensurate peaks in the charge correlation
at ${\bf q}$=$(2\pi{\rm x},2\pi{\rm x})$.
The spin sector is also nontrivial (Fig.~3(c)) and the state is
an insulator according to the DOS (Fig.~3(d)) \cite{comment2}.

In summary, the extended phase diagrams of manganites for x=0 and
x$>$0 have been discussed.
A novel E-AFM phase, stabilized at x=0 in the region of weak- and
intermediate-couplings,
is adjacent to both FM metallic and A-AFM states.
The competition between E-AFM insulating and FM metallic phases
suggests the possibility of CMR effects even in undoped manganites.
Several features of the E- to A-AFM transition, 
at least at the qualitative level, agrees with currently available
experimental results.
For the doped case, a microscopically inhomogeneous
C$_{\rm x}$E$_{\rm 1-x}$-AFM state is predicted.
This state may contribute to the phase separation tendencies widely
observed experimentally in Mn-oxides for 0$\alt$x$\alt$0.5.
The discovery of these many ``hidden'' interesting phases in real 
manganites -- undoped and doped --
should be actively pursued experimentally. This will open a new sub-branch
of investigations in the already much interesting context of Mn oxides.

The authors thank T. Kimura and K. Ueda for discussions.
T.H. has been supported by a Grant-in-Aid for Scientific Research from
the Ministry of Education, Culture, Sports, Science, and Technology of Japan.
E. D. is supported by the NSF grant DMR-0122523.



\begin{references}
\bibitem{comment3}
Although x=0 corresponds to one $e_g$ electron per site which naively
may suggest insulating behavior, there are two active
orbitals and the mean density per state is 1/2. Note also that
previous work \cite{Dagotto1} has shown that a large Hund coupling
is equivalent to a large Hubbard repulsion.
\bibitem{note:Coulomb1}
Since JT distortions suppress double occupancy as short-range Coulomb
interaction does, a large $\lambda$ is quite fine to mimic the
physics of manganites, as shown in \cite{Dagotto1}.
\bibitem{note:CE}
A similar mechanism has been discussed by some of 
the authors for the CE-type phase at x=0.5 \cite{Hotta4}.
\bibitem{note:TcandTN}
While finite clusters do not show true thermodynamic singularities,
a rapid increase in the strength of correlations is observed
at fairly well defined temperatures, here referred to as $T_{\rm N}$
and $T_{\rm C}$ for simplicity.
Previous computational experience shows that these
estimations are fairly accurate upon increasing lattice sizes
\cite{Dagotto1}.
\bibitem{note:opt}
Both spin directions and oxygen positions are optimized.
\bibitem{note:FMOO}
Note that the FM/OO state of Fig.~2(b) is another phase not previously
observed experimentally, here predicted to exist for low-$J_{\rm AF}$
manganites at x=0.
\bibitem{note:Coulomb2}
As in any striped state, further work should be carried out to
check the stability of the structure against long-range Coulomb
interactions.
\end{references}
\end{document}